# Quantum efficiency enhancement of mid infrared photodetectors with photon trapping micro-structures

Ekaterina Ponizovskaya Devine,[1] Hilal Cansizoglu,[1] Yang Gao,[1] Soroush Ghandiparsi,[1] Cesar Perez[1], Hasina H. Mamtaz,[1] H. Rabiee[1], and M. Saif Islam[1]

**Affiliations:**

[1]Electrical and Computer Engineering, University of California, Davis, Davis, CA 95618.

**Abstract**
The study proposes to use the photon trapping micro-structures to enhance quantum efficiency of the mid infrared photodetectors. The nanostructure that is consist of micro holes reduces reflection and bends the near normally incident light into the lateral modes in the absorbing layer.

**Summery**

The recent study show interest in mid infrared photodetectors (MID) for room temperature [1,2]. The new 2D materials such a b-AsP or graphene can show significant absorption for infrared wavelength. Our study proposes the photon trapping nanostructure that enhances performance of the traditional HgCdTe [3] photodetectors as well enhance the coupling of the infrared radiation with the new 2D materials. Recent theoretical and experimental studies [4] showed that micro-holes nanostructures enhanced quantum efficiency (QE) for fast Si photodetectors for data communication. We optimized the micro hole parameters for the MID. Our simulations show that the QE can be increased and the operation region can be widened toward the region where the absorption of the material is relatively small. The QE enhancement also can increase the MID performance at high temperature.

This study presents Finite Difference Time Domain (FDTD) simulation shows that the lateral modes that propagates along the PDs surface are responsible for the infrared radiation trapping. Optimizing the holes structure, we found that the holes comparable to the wavelength provide the better bending of the infrared into the lateral modes. The designed PDs have thin (30-50nm) b-AsP layer on a Si substrate covered protective layer with 2D periodic micro-holes. The nanoholes view from the top and cross-sections are in the x-z plane (perpendicular to the surface) are shown in Fig. 1a,b,c. The PD can also use HgCdTe with holes in the absorbing material. In this case a thinner layer of HgCdTe can be used. We considered Si or $SiO_2$ protective covers. The hole shape varied from cylindrical to tapered cylindrical and inverted pyramids (Fig.1b, c). The holes array supports a set of modes with wave vectors in z direction, $k_z$ and $k_c$ in x-y plane (parallel to the surface). Each hole couples the vertically incident infrared radiation into the x-y plane. The lateral modes with larger $k_c$ can have full reflection from the bottom of the PD and they can form modes guided in the x-y plane in b-AsP layer. The lateral modes [4,5] in xy plane causing higher absorption. The FDTD simulation for the inverted pyramids holes in $SiO_2$ cover with period 2000nm and diameter of 1700nm (Fig.1d green curve) shows some absorption increase over flat cover (Fig.1d, black curve). The parameters for the b-AsP layer were used based on experiment [1] that showed about 10-15% QE. The same holes in Si cover produced much better absorption enhancement (Fig.1d, red curve). The lateral modes are responsible for the QE enhancement [5]. The longer the light propagate into b-As layer and the better it confined in it the better is the absorption for a wider range of the wavelength. The field distribution around the cylindrical holes (Fig.1e) shows the lateral modes between the holes. The tapered or inverter pyramids shape had smaller reflection due to the effect is similar to the Lambertian reflector that helps to trap infrared radiation [5]. The micro-holes also provide better absorption in wider region.

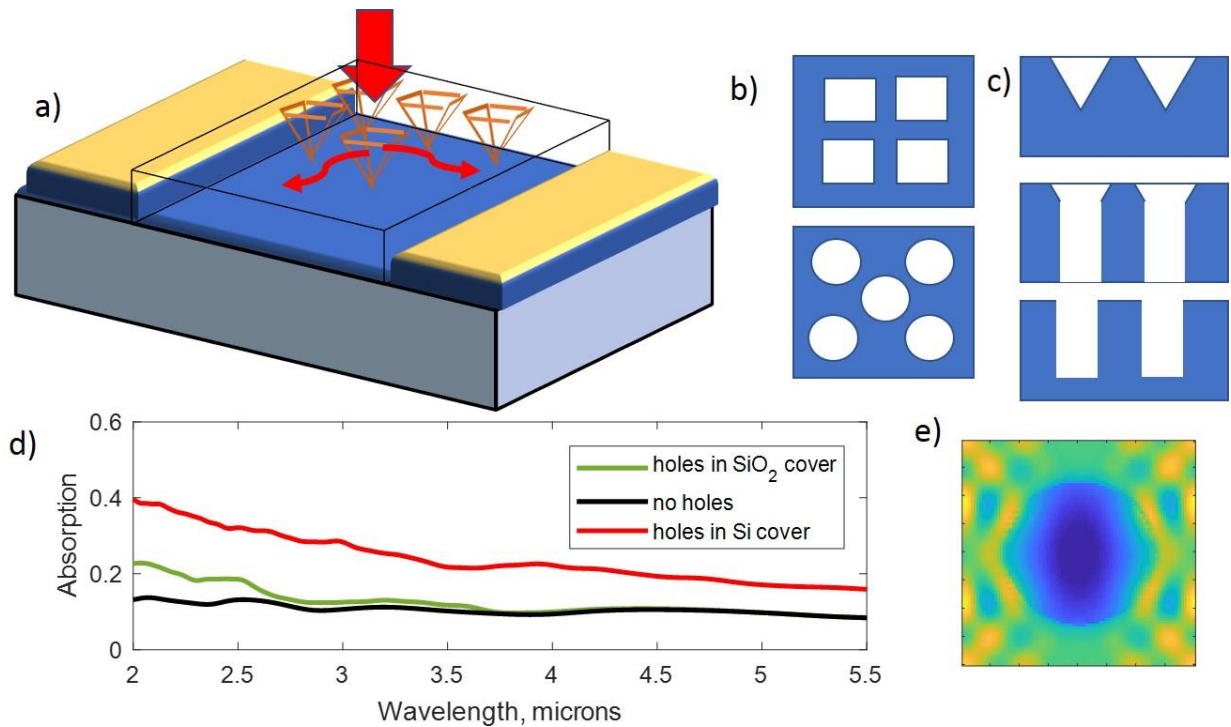

Fig.1. Absorption enhancement: a) PD holes in the cover; b) holes shape top view; c) holes shape side view; d) absorption with flat silicon oxide cover (black), with holes in silicon oxide cover (green), with holes in Si cover (red), e) typical field distribution around cylindrical holes.

The FDTD simulations showed that inverted pyramids holes with period 2000nm and 1700nm in Si cover produced the best coupling with absorbing b-AsP layer and can significantly increase QE.

The proposed nanostructure reduces reflection and bends the near normally incident light into the lateral modes, that can be used to increase QE of 2D materials like graphene or p-AsP. The PD can also use HgCdTe or PdSe with holes in the absorbing material. In this case the absorbing layer can be smaller and the dark current can be reduced and the PD could be used at higher temperature.

## Literature

1. Mingsheng Long, Anyuan Gao,1Peng Wang,Hui Xia, Claudia Ott, Chen Pan, Yajun Fu, Erfu Liu, Xiaoshuang Chen, Wei Lu, Tom Nilges, Jianbin Xu, Xiaomu Wang, Weida Hu, Feng Miao. Room temperature high-detectivity mid-infrared photodetectors based on black arsenic phosphorus. *Science Advances*  30 Jun 2017:Vol. 3, no. 6,e1700589 DOI: 10.1126/sciadv.1700589
2. Chao-Hui Yeh, Zheng-Yong Liang, Yung-Chang Lin, Tien-Lin Wu, Ta Fan, Yu-Cheng Chu, Chun-Hao Ma, Yu-Chen Liu, Ying-Hao Chu, Kazutomo Suenaga, and Po-Wen Chiu Scalable van der Waals Heterojunctions for High-Performance Photodetectors *ACS Appl. Mater. Interfaces*, 2017, *9* (41), pp 36181–36188 DOI: 10.1021/acsami.7b10892
3. A Rogalski, HgCdTe infrared detector material: history, statusand outlook,  Rep. Prog. Phys. **68** (2005) 2267–23362.
4. Yang Gao, Hilal Cansizoglu,Kazim G. Polat,, Soroush Ghandiparsi, Ahmet Kaya, Hasina H. Mamtaz[1], Ahmet S. Mayet[1], Yinan Wang[1], Xinzhi Zhang[1], Toshishige Yamada[2, 3], Ekaterina Ponizovskaya Devine ORCID: orcid.org/0000-0003-0768-5981[3], Aly F. Elrefaie[1, 3], Shih-Yuan Wang[3] & M. Saif Islam, Photon Trapping Microstructures Enable High-Speed High Efficiency Silicon Photodiodes,  Nature Photonocis, DOI: 10.1038/NPHOTON.2017.37  (2017) 5.E.Ponizovskaya Devine et al, Optimization of light trapping micro-hole structure for high-speed high-efficiency silicon photodiodes Conference: IEEE Photonics Conference (IPC), Orlando, FL, Volume: 587-588 (2017)